\documentclass{rjparticle}
\usepackage{graphicx}

\newcommand{\miktex}{\hbox{Mik\kern-.15em\TeX}}

\newcommand{\be}{\begin{equation}}
\newcommand{\ee}{\end{equation}}
\newcommand{\lmaxtune}{\ell_{\mathrm{max, tune}}}
\newcommand{\unit}[1]{\ensuremath{\,\mathrm{#1}}}

\title{The Oddly Quiet Universe: How the CMB challenges cosmology's standard model} 
\author[1]{Glenn D. Starkman}
\author[1]{Craig J. Copi}
\author[2]{Dragan Huterer}
\author[3]{Dominik Schwarz}
\affil[1]{Department of  Physics/CERCA/ISO,\\Case Western Reserve University\\Cleveland, OH 44106-7079 USA
\\Email:{\em gds6@case.edu, cjc5@case.edu}}
\affil[2]{Department of  Physics\\University of Michigan\\ Ann Arbor, MI 48109-1040 USA
\\Email:{\em huterer@umich.edu}}
\affil[2]{Fakult\"at f\"ur Physik\\Universit\"at Bielefeld \\
Postfach 100131, D-33501 Bielefeld, Germany\\Email:{\em dschwarz@physik.uni-bielefeld.de }}
\keywords{cosmology, cmb, anomalies}
\pacs{01.30.-y, 01.30.Ww, 01.30.Xx}

\begin{document}
\maketitle
\begin{abstract}
We discuss selected large-scale anomalies in the maps of temperature anisotropies in the cosmic microwave background. 
Specfically, these include alignments of the largest modes of CMB anisotropy with one another and with the 
geometry and direction of motion of the Solar System, and the unexpected absence of two-point angular corellations especially
outside the region of the sky most contaminated by the Galaxy.
We discuss these findings in relation to expectations from standard inflationary cosmology. 
This paper is adapted from a talk given by one of us (GDS) at the SEENET-2011 meeting  in August 2011 on the Serbian bank of the 
Danube River.
\end{abstract}

When we look up at the sky with microwave eyes, we see a uniform glow, such as would be emitted by a black body of just under $3^\circ$K.
This is of course the Cosmic Microwave Blackbody  (CMB) radiation originating from when the expansion of the Universe had cooled and diluted
the primordial plasma to the point where neutral atoms were finally long-lived, and the Universe thus became transparent.  
This CMB was first measured in the 1960s by Penzias and Wilson. Its discovery was key in establishing the Big Bang theory of cosmology's primacy over
its  rival, the Steady State theory.   In the 1970s, a dipolar deviation from the isotropy of that signal was first measured,
reflecting predominantly, it is believed, a Doppler-induced signal due to the Solar System's motion through the rest frame defined by the CMB.
It took until the 1994 for the COsmic Background Explorer (COBE) satellite's Differential Microwave Radiometer (DMR) to finally measure and make a 
full-sky map of intrinsic anisotropies in the radiation at the level of a few tens of $\mu$K.    These anisotropies represent principally fluctuations in the temperature/density/gravitational potential at the time of the photons' last scattering, also known as the epoch of recombination.  They are thus the seeds of cosmic structures that would grow to become galaxies, clusters of galaxies, and, eventually more complicated non-linear inhomogeneites such as us.

While many experiments in the 1990s and 2000s were able to improve on aspects of the COBE DMR measurements -- increasing the S/N and the angular resolution over small
patches -- it was only with the publication of the first results from the Wilkinson Microwave Anisotropy Probe  (WMAP) in 2003 that there was an improved full-sky map of the CMB temperature fluctuations.   One might well wonder how one can make a full-sky map of the CMB when the Galaxy intervenes and is brighter than the CMB over the part
of the sky close to the Galactic plane.   Indeed, in COBE maps released to the public, the Galactic plane is immediately evident.  Meanwhile publicity maps from WMAP show no Galactic plane.  This is because with improved S/N and more frequency bands, WMAP is able to produce a map out of linear combinations of the individual waveband maps in which
the Galactic foregrounds should be, and at least appear to be, subdominant.   This is the so-called Internal Linear Combinations (ILC) map.  Similar maps have been made by independent groups of researchers and are quite close to the ILC.   Nevertheless, one should not be surprised should one discover remnant Galactic contamination in such
maps, and for most purposes CMB analysis is done on partial skies in which the Galactic Plane and other contaminated regions (such as those surrounding bright
point sources) have been removed.

Although the WMAP temperature maps most obviously improve our ability to infer the small-angular-scale properties of the CMB, they also, 
especially because of their increased signal to noise, significantly increase our knowledge of the large scale properties of the CMB.  It is 
on these large scale properties that this paper focuses.   In the rest of this paper we will investigate the large angle properties of the CMB.
We will discover that if one uses the full-sky ILC map then one finds very odd correlations in the map, that correlate unexpectedly to the Solar System.
(Unexpectedly, because we were led to believe that any residual contamination in the map would be connected to the Galactic foregrounds.) 
By masking out  the region of the sky behind the Galactic plane, we will then discover that the sky appears to have remarkably little correlation between
pairs of points separated by more than $60^\circ$, at a level that is statistically very surprising.  Looking into this anomaly more deeply we will
find that it remains robust through all seven years of published WMAP data, and furthermore that it is very difficult to explain within the context of
the canonical  Inflationary Lambda Cold Dark Model model of cosmology.  

\section{Multipole Decomposition}
\begin{figure}[h!tb]
\centering
\includegraphics[width=0.8\textwidth]{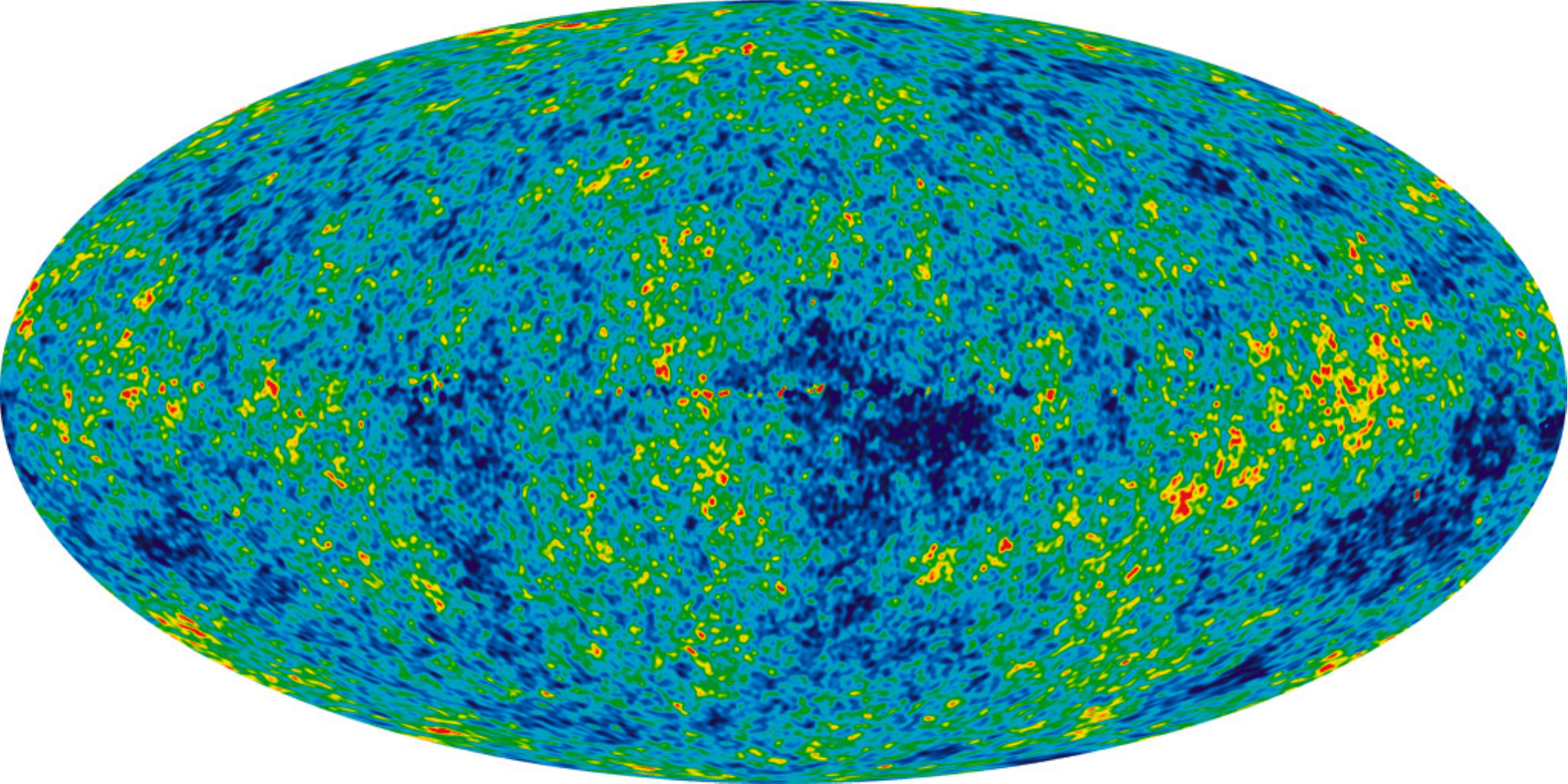}
\caption{The Internal Linear Combination Map is a weighted linear combination of the five WMAP frequency maps. The weights are computed using criteria which minimize the Galactic foreground contribution to the sky signal. The resultant map provides a low-contamination image of the CMB anisotropy. Courtesy of the WMAP Science Team.}
\label{fig:ilc_7yr}
\end{figure}

When confronted by WMAP's all sky map of the CMB temperature fluctuations (Figure \ref{fig:ilc_7yr}), the immediate response of  a cosmologist  is to expand the map in spherical harmonics:
\be
\frac{\Delta T}{T}(\theta,\phi) \equiv \sum_{\ell=2}^{\infty} \sum_{m=-\ell}^\ell a_{\ell m} Y_{\ell m}(\theta,\phi) \,,
\ee
the monopole ($\ell=0$) and dipole  ($\ell=1$)  having been subtracted.   This expansion is so automatic because the inflationary model tells us that the $a_{\ell m}$
are (realizations of) Gaussian random variables of zero mean.  (Or nearly so.  Non-linear effects can induce small, but in-principle measurable non-Gaussianity.) 
The $a_{\ell m}$ are therefore the most convenient physical variables for comparing observations with theory.
Moreover, in the standard Lambda Cold Dark Matter ($\Lambda$CDM) model, the Universe is statistically isotropic, so that the expectation values of $a_{\ell m}$
obey the relation
\be
 \label{eqn:Sl}
\langle a^\star_{\ell m}  a_{\ell^\prime m^\prime} \rangle = C_\ell \delta_{\ell\ell^\prime} \delta_{m m^\prime}.
\ee
This means that, in the standard theory, the only thing worth measuring is $C_\ell$.   
 The variances, $C_\ell$,  of the underlying Gaussian variables, $a_{\ell m}$, are also the expected values of the measured angular power spectrum, 
 \be
 \label{eqn:pseudoCl}
 C_\ell=\frac{1}{2\ell+1}\sum_{m=-\ell}^\ell \vert a_{\ell m}\vert^2 \,.
 \ee
 We shall very sloppily use the same symbol for both the $C_\ell$ appearing in equation \ref{eqn:Sl} and the $C_\ell$ appearing in equation \ref{eqn:pseudoCl},
 even though the former is a property of the underlying statistical distribution of which the sky is a realization, and the latter is a property of the actual sky.
 More pedagogically careful treatments are readily found in the literature, but this level of sloppiness is standard.

\begin{figure}[h!tb]
\centering
\includegraphics[width=0.8\textwidth]{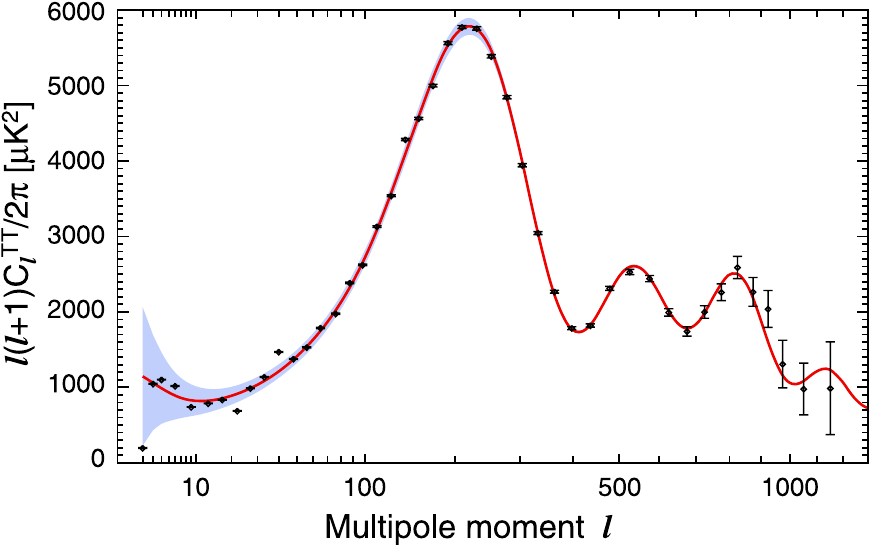}
\caption{The TT Angular power spectrum.  The points are the 7-year temperature (TT) power spectrum from WMAP. The curve is the $\Lambda$CDM model best fit to the 7-year WMAP data. The plotted errors include instrument noise, but not the small, correlated contribution due to beam and point source subtraction uncertainty. The gray band represents cosmic variance. Figure is from \cite{Larson:2010gs} courtesy of the WMAP Science Team.}
\label{fig:aps_wmap7}
\end{figure}

The WMAP angular temperature-temperature (TT) power spectrum is shown in Figure \ref{fig:aps_wmap7}.
(To be specific, this is the power spectrum produced by the WMAP Science Team using the first  seven years of data.)
Fitting cosmological parameters to the inflationary $\Lambda$CDM model allows us  to infer important properties of the
Universe within the context of that canonical model.   In particular, given our interest in the largest scale properties of the Universe,
we learn that the Universe has a geometry that is indistinguishable from flat.  (Meaning that no curvature, either positive or negative
can be discerned.)   This can be seen approximately by the location of the first peak in the power spectrum, which is in the bin centered around $\ell=91$.
(The first clear detection of the first peak was made by the TOCO experiment \cite{Torbet:1999sg}, and then by the Boomerang collaboration \cite{deBernardis:2000gy}
who were the first to conclude  that the Universe is the therefore close to flat.)

Traditionally, the geometry of the Universe was the only ultra-large scale property of the Universe that one needed to measure;
which would lead us to ask whether there is anything else interesting to learn about the Universe on largest scales.   As we shall
see, the data suggests that there is.  We might begin to suspect that this is the case by looking at the angular power spectrum
and observing that the value of the quadrupole $C_2$ is anomalously small -- well outside the grey cosmic variance band.  Just how
unlikely this is has been a matter of extensive, but rather uninformative, debate.  Uninformative, because it is really not the
smallness of the quadrupole that we shall conclude is really strange about the large-angle CMB sky.   Nevertheless, it  was one
of the motivating factors behind various investigators explorations of the low-$\ell$, or large angle, properties of the CMB.

As this is not a review, we shall not attempt to be exhaustive or even comprehensive in our exploration of large angle CMB anomalies.  
There are reviews of the subject which attempt to be so.  Two very different viewpoints are offered by the WMAP Science Team itself
in \cite{Bennett:2010jb}, and this collaboration in \cite{Copi:2010na}.   We shall instead focus here on two
results that we have highlighted which reflect the corner that the large angle CMB seems to paint us into.  
This is just a small part of the big picture, and we apologize to our many colleagues whose fine (!) work we do not cite here.  

\begin{figure}[h!tb]
\centering
\includegraphics[width=0.8\textwidth]{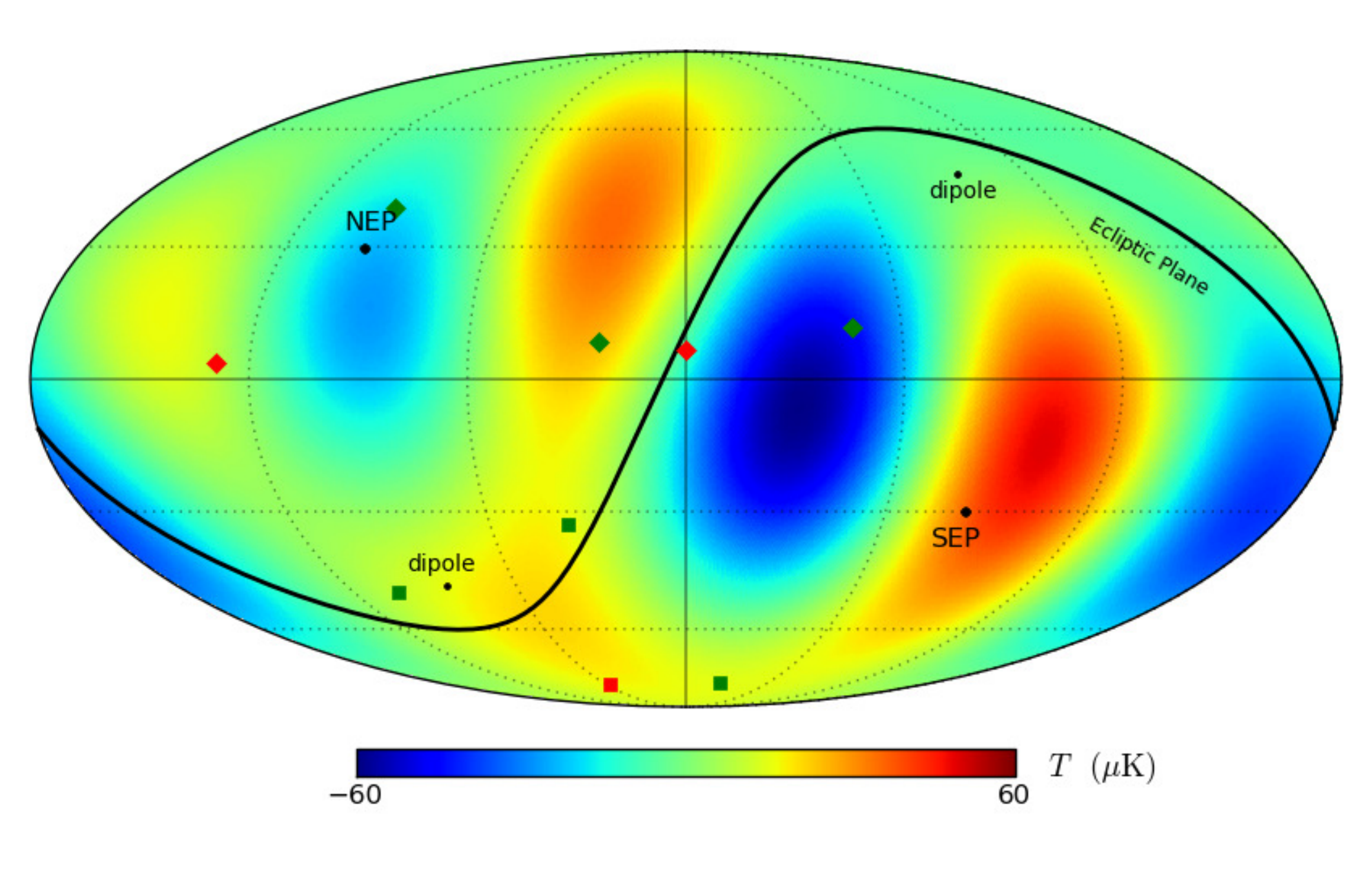}
\caption{Quadrupole plus octopole anisotropy of the WMAP sky map in Galactic coordinates, shown with the ecliptic plane and the cosmological dipole. 
Included are the multipole vectors (solid diamonds); two for the quadrupole (red diamonds) and three for the octopole (green diamonds). 
We also show the four normals (solid squares) to the planes defined by the vectors that describe the quadrupole and octopole temperature anisotropy.
Figure from \cite{Copi:2010na}.}
\label{fig:ilc3_2_3}
\end{figure}

First, we shall work with the full sky CMB as represented by the ILC.   (It doesn't seem to much matter which year ILC one uses,
or whether one instead uses one of the other full sky maps, so we shall just stick with the ILC map which was produced using three years of WMAP data.)
We shall look only at the two lowest interesting monopoles $\ell=2,3$.  
In Figure \ref{fig:ilc3_2_3}, 
we plot the quadrupole $\ell=2$ plus octopole anisotropy of the ILC sky map in Galactic coordinates. 
(The $\ell$th multipole is just $\sum_{m=-\ell}^\ell a_{\ell m} Y_{\ell m}$.)
Various other environmental quantities are shown -- the plane of the ecliptic (the plane of the Solar System) together with the north and south
ecliptic poles (the normal directions to that plane), and  the cosmological dipole direction (and its antipode).

We have also included the multipole vectors that describe the quadrupole and octopole.
Multipole vectors are the analogs for $\ell>1$ of the dipole vector.  
We normally think of a pure dipolar real function $f_1(\theta,\phi)$ in terms of a vector ${\vec d}$,
$f(\theta,\phi) = {\vec d}\cdot {\hat r}$ instead of as a sum of spherical harmonics.   
(Here $\hat r$ is the unit coordinate vector, ${\hat r} = (\sin\theta\cos\phi,\sin\theta\sin\phi,\cos\theta)$.)   
The magnitude of the dipole is $d\equiv \vert \vec d\vert$ and its direction  is the unit vector ${\hat d} \equiv {\vec d}/d$.
The dipole strength $d$ plus the two degrees of freedom of $\hat d$ replace the three real coefficients $\mathrm{Re}(a_{1 1})$, $\mathrm{Im}(a_{1 1})$ and $a_{1 0}$ 
of the spherical harmonic expansion. 
(Since the function $f$ is real, $a_{10}$ is real, and $a_{1 -1}$ is determined by $a_{11}$.) 

Similarly, we can replace the five real degrees of freedom of the quadrupole ($a_{2m}$ as constrained by reality conditions) 
with two unit vectors ${\hat u}^{2,1}$ and ${\hat u}^{2,2}$ and a single scalar $A^{(2)}$. 
 (This is because an angular momentum $2$ object can be obtained from  the product of two angular momentum one objects.)
These unit vectors are the multipole vectors of the quadrupole.
Similarly the seven real degrees of freedom of the octopole ($a_{3m}$) can be replaced by three unit vectors ${\hat u}^{3,1}$, ${\hat u}^{3,2}$ and ${\hat u}^{3,3}$
and a single scalar $A^{(3)}$.     (This multipole vector representation appeared as early as \cite{Maxwell}.)

The multipole vectors of the quadrupole appear in Figure \ref{fig:ilc3_2_3} as red diamonds, those of the octopole as
green diamonds.  They are plotted in both northern and southern hemispheres because they are defined only up to a sign that can be absorbed into $A^{(i)}$.
We also plot the normals to the plane defined by the two quadrupole multipole vectors  ${\hat n}^{(2,1,2)} \parallel ({\hat u}^{2,1}\times {\hat u}^{2,2})$ as
a red square (again in both hemispheres), and as green squares the normals to the three planes defined by the three multipole vectors of octopole.
Note that the normals cluster together on the sky, implying that quadropole plane and the three octopole planes are nearly aligned.
Moreover, the normals are near the ecliptic plane, implying that not only are these four planes aligned but the  are nearly perpendicular to the ecliptic.
Furthermore the normals are near the dipole, meaning that the planes are not just aligned and perpendicular to the ecliptic
but oriented perpendicular to Solar System's motion through the Universe.   
Finally, as one can see from Figure \ref{fig:ilc3_2_3}, the great circle of the ecliptic plane (black curve), very carefully separates the
strong extrema to its south from the weaker extrema to its north.
The precise statistical significance of these correlations, first discussed in \cite{Copi:2003kt} 
(although the alignment of the octopole and quadrupole with one another was first pointed out in \cite{deOliveiraCosta:2003pu}), 
depends on how one calculates them.   

However one does the statistical analysis, 
these apparent correlations with the Solar System geometry are puzzling.
They do not seem to reflect the Galactic contamination that we might have expected from residual foreground contamination in the ILC map.
Indeed there are a number of challenges to explaining these results.   For
one, the observed quadrupole and octopole are aligned (appear as $Y_{\ell \vert\ell\vert}$
in a frame with the $z$-axis along the common or average axis of the four
planes), and not as a $Y_{\ell 0}$ (in any frame).   This makes it
difficult to explain them in terms of some localized effect on the sky.   Also, the quadrupole is much smaller than the octopole, which
means that perturbative explanations in terms of small vectors or gradients are challenging.   The best one can say is that
these full-sky solar-system correlations remain unexplained.  
The Planck experiment will hopefully shed new light on these mysteries.

\section{Angular Correlation Function}

We now wish to leave the harmonic domain and look instead at the real space two-point correlation function of the CMB temperature map,
\be
C(\theta) = \overline{ T(\hat n_1)  T(\hat n_2) }\vert_{\hat n_1\cdot\hat n_2=\cos\theta} \,,
\ee
where the overbar indicates an average over all pairs of directions on the sky separated by the angle $\theta$.

While it was once traditional to calculate the angular two-point correlation function, this has mostly fallen out of favor.
Partly this is because of the lore that $C(\theta)$ contains exactly the same information as the angular power spectrum $C_\ell$,
\be
\label{eqn:CthetaLegendre}
C(\theta) = \sum_\ell \frac{2\ell+1}{4\pi} C_\ell P_\ell(\cos\theta) \,.
\ee
However, this relationship between observed quantities holds only when evaluated on the full sky.
If we impose a Galaxy cut on the map before evaluating the $C_\ell$ or $C(\theta)$, as is typically done,
then the relationship fails.
Also, it holds in the statistical ensemble only if the assumption/prediction of statistical isotropy is correct.
It may be, but it certainly should be tested.  
Finally, there is a reason why we often look at both a function and its Fourier  (or other appropriate) transform -- 
features that are hidden in one sometimes become much more evident in the other.

\begin{figure}[h!tb]
\centering
\includegraphics[width=0.8\textwidth]{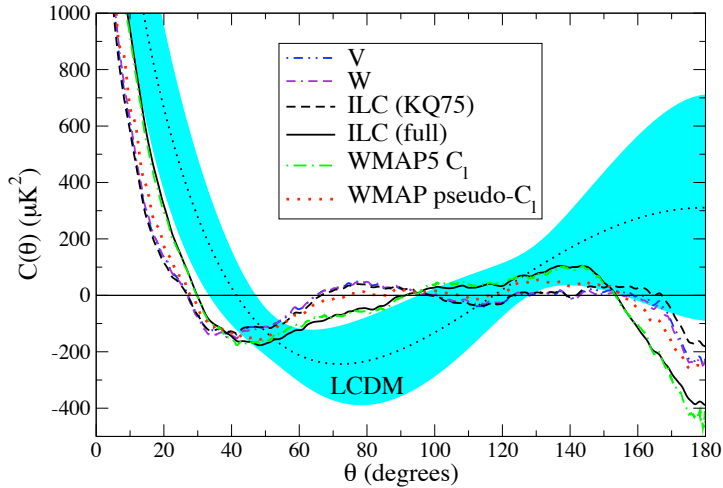}
\caption{The angular two-point angular correlation function from the WMAP 5 year results. $C(\theta)$ is plotted for maps with their monopole, dipole and
Doppler  quadrupole subtracted.
 The V-band (dashed-dotted-dotted line), W-band (dashed-dashed-dotted line), ILC (KQ75, dashed line) have had the KQ75 mask applied. 
 The full-sky ILC result (solid line) is also shown. Also plotted are $C(\theta)$ from the WMAP maximum likelihood $C_\ell$ (dotted-dashed line), 
 the WMAP pseudo-$C_\ell$ (dotted line) and the best-fit $\Lambda$CDM $C_\ell$. 
 The shaded region is the one sigma cosmic variance bound on the standard $\Lambda$CDM theory. Figure from \cite{Copi:2010na}.}
\label{fig:ctheta_5yr}
\end{figure}

In figure \ref{fig:ctheta_5yr} we plot various versions of the two point angular correlation function from the WMAP 5-year data.
The smooth dotted line with the blue band around it is the $C(\theta)$ that would be obtained from equation \ref{eqn:CthetaLegendre}
using the angular power spectrum $C_\ell$ predicted by the best-fit $\Lambda$CDM model.   That blue band is the one-sigma
cosmic variance band.  In other words, if we vary each of the $C_\ell$ inside the one-sigma cosmic variance range around
its expected value, then the $C(\theta)$ obtained will remain entirely within the blue region.

While the smooth blue band is the theoretical expectation, all of the jagged curves are obtained from the data in one way or another.
Our first observation is that none of those data curves look like the theory curve.  They do not remain inside the blue band.   
However, one cannot really compare curves by eye  because the different points on the $C(\theta)$ curve are highly
correlated; we must devise some statistical measure of the their difference.  Nevertheless
we shall not focus on the difference between the data curves and the theory curve.   

What is actually more striking (and more significant) is that {\bf all} of the $C(\theta)$ curves that are calculated excluding the region inside the Galactic plane
remain remarkably close to $C(\theta)=0$ over a very wide range of $\theta$, from about 60 to 170 degrees. We can quantify this in terms of a statistic
first suggested by the WMAP Science Team \cite{Spergel:2006hy}:
\be
\label{eqn:Shalf}
S_{1/2} \equiv \int_{-1}^{1/2} d\cos\theta \left[ C(\theta) \right]^2
\ee
We have shown that the p-value of $S_{1/2}$ for the five year ILC outside the KQ75 Galactic cut is a remarkably tiny $0.025\%$.

We see in figure \ref{fig:ctheta_5yr} that the six data-derived curves divide neatly into two classes.  Four that hug the $C(\theta)=0$ axis
closely on scales $\theta\geq60^\circ$, and two that do not.  All four of the zero-huggers are calculated by taking the sky average in (\ref{eqn:CthetaLegendre})
{\em only} over  pairs of points {\em both} of which are outside the Galaxy (as defined by the WMAP Science Team's KQ75 Galaxy cut).  
In other words they involve direct calculations of $C(\theta)$ over the parts of the sky that are to be trusted.
Instructively, the zero-huggers  include all three individual cut-sky (ie. KQ75-masked) waveband maps -- Q, V, W.    It is no surprise that the cut-sky ILC
rounds out the group of four, since the ILC is just a linear combination of the individual band maps.   

The other two data-derived $C(\theta)$ curves {\em also} agree with one another, but not with the zero-huggers (nor with the theory curve or error band).
They include the full-sky ILC, i.e. the ILC with no Galaxy cut, and the curve generated by substituting the WMAP-reported angular power spectrum $C_\ell$ 
(derived using a Maximum Likelihood Estimator -- MLE) into equation (\ref{eqn:CthetaLegendre}).   
Considering  the full-sky ILC vs. the cut-sky ILC, we learn that essentially all of the large-angle ($\theta > 60^\circ$) 
angular correlation on the sky is due to pairs of points at least one of which is {\em inside} the Galaxy, ie. inside the part of the sky that we don't trust.

The difference between the MLE curve and the cut-sky curves is less straightforward, since, as far as we can tell, the WMAP-supplied MLE $C_\ell$ also derive from cut sky data.
However,  the discrepancy may well trace to the same cause as identified in \cite{Copi:2011pe}  which addresses the argument of \cite{Efstathiou:2009di}
that one should first reconstruct the full sky from the cut sky and then calculate the full-sky two-point correlation function. 
 Reference \cite{Copi:2011pe} shows that  in practice the reconstruction  of $C_\ell$ is biased due to leakage of
information from the region obscured by foregrounds to the region used for the reconstruction.  
 This leakage comes because of the need to smooth the map before reconstructing.  
In the region oustide but near the cut, he smoothing incorporates data from inside the cut.
Since the cut is imposed because the data inside it is unreliable, one must decide what to substitute for that data.
The results then depend on the choice of how to fill the cut.   Not surprisingly, the problem is largest when the particular
$C_\ell$ one is reconstructing is anomalously small (compared to $C_\ell$ of nearby $\ell$).  Of course, this is 
precisely the case for the quadrupole.      Reference \cite{Copi:2011pe}  did not extend the analysis to $C(\theta)$, 
or more specifically $S_{1/2}$, but one may reasonably suspect  that the reconstruction is similarly (maybe even particularly) poor
at maintaining $C(\theta)=0$ since that property will not be maintained by most choices of how to fill the cut region.

\begin{figure}[h!tb]
\centering
\includegraphics[width=0.8\textwidth]{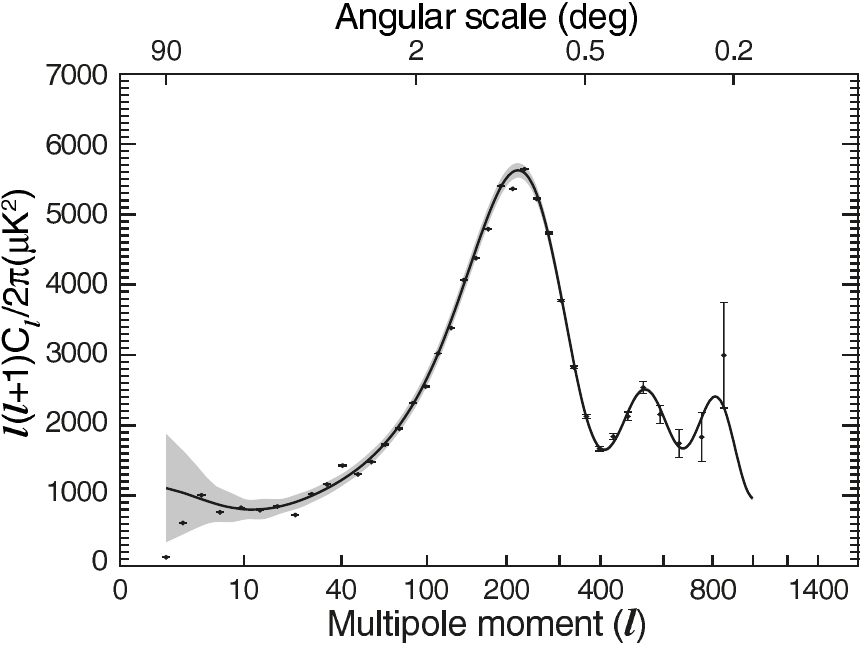}
\caption{The WMAP first year TT Angular power spectrum.  
 The data are plotted with $1\sigma$ measurement errors. 
 The solid line shows the best-fit $\Lambda$CDM model from \cite{Spergel:2003cb}. 
 The gray band around the model is the $1\sigma$ uncertainty due to cosmic variance on the cut sky. 
 For this plot, both the model and the error band have been binned with the same boundaries as the data, but they have been plotted as a splined curve to guide the eye. 
 On the scale of this plot the unbinned model curve would be virtually indistinguishable from the binned curve except in the vicinity of the third peak.
Figure is from\cite{Hinshaw:2003ex} courtesy of the WMAP Science Team.}
\label{fig:aps_wmap1}
\end{figure}

The absence of two-point correlation on large scales is a much larger problem than the smallness of $C_2$ that first led the community to worry 
about the low-$\ell$ CMB sky.   There are two ways to have a sky with low $C(\theta>60^\circ)$.  One is to have all the $C_\ell$ small  (or at least
all the $C_\ell$ up to some sufficiently high $\ell$).   This is not our observed sky.   
(For one thing, such a sky would almost certainly also have a low $C(\theta<60^\circ)$.)  
Looking at figure \ref{fig:aps_wmap7} we see that most of  the $C_\ell$ are comparable to their theoretical values.   
But recall that the WMAP science team used an MLE method to infer the low-$\ell$ $C_\ell$.   This was not the
case in their first-year data release, so it is better to look at figure \ref{fig:aps_wmap1} to confirm that only $C_2$
is particularly small.     

The other way to get $C(\theta\geq60^\circ)\simeq0$ (but not  $C(\theta\leq60^\circ)\simeq0$) is for the low-$\ell$ $C_\ell$ to have
a particular relationship to one another.  What relationship?  The one obtained by expanding the observed $C(\theta)$ in the
Legendre series  of equation \ref{eqn:CthetaLegendre}.   However, these relationships are delicate and imply that
the $C_\ell$ must be correlated with each other.  \begin{table*}
  \begin{minipage}{4.5in}
    \caption{$S_{1/2}$ (in $(\mu\mathrm{K})^4$) obtained by minimizing with respect to $C_\ell$ 
    (for $\ell$ in the range $2\leq \ell\leq \lmaxtune$ and fixing $C_\ell$ with $\ell>\lmaxtune$). 
    We show the statistic for the best-fit theory and for WMAP, as a function of the cutoff multipole $\lmaxtune$.  
    Also shown  is the 95 per cent confidence region of the minimized $S_{1/2}$ derived from chain 1 of the WMAP MCMC parameter fit.  
    The bottom row gives the measured value of $S_{1/2}$ outside the cut -- $1152\unit{(\mu K)^4}$.   Table is taken from \cite{Copi:2008hw}.}
    \label{tab:minS12}
    \begin{tabular}{lccccccc}
      \hline
      $C_\ell$ & \multicolumn{7}{c}{Maximum tuned multipole,  $\lmaxtune$} \\ \cline{2-8}
      Source & $2$ & $3$ & $4$ & $5$ & $6$ & $7$ & $8$ \\ \hline
      Theory & $7624$ & $922$ & $118$ & $23$ & $7$ & $3$ & $0.7$ \\
      Theory 95\% & $6100$--$12300$ & $750$--$1500$ & $100$--$200$ &
      $20$--$40$ & $7$--$14$ & $3$--$6$ & $1$--$3$ \\
      WMAP & $8290$ & $2530$ & $2280$ & $800$ & $350$ & $150$ & $130$ \\
      \hline
        ILC5 (KQ75) & \multicolumn{7}{c}{$1152$}\\
        \hline
    \end{tabular}
  \end{minipage}
\end{table*}
Table \ref{tab:minS12}, taken from \cite{Copi:2008hw}, shows how, given the higher $\ell$ 
$C_\ell$ observed by WMAP, it is necessary to tune the contributions to $C(\theta)$ from $C_2$, $C_3$, $C_4$ and $C_5$ 
against those from $C_6$ and above in order to get $S_{1/2}$ to be as low as it is ($1152\unit{(\mu K)^4}$).
By contrast, in the best fit theory it would have been enough to tune just $C_2$ and $C_3$.

It  is extremely difficult to arrange for the $C_\ell$ to have particular {\em relative} values in the context of the standard inflationary model,
because  the $a_{\ell m}$ are {\em independent} Gaussian random variables.   Thus, even if we were able to adjust the theory so that
the expected values of the $C_\ell$ ({\it i.e.} of $(2\ell+1)^{-1}\sum_m \vert a_{\ell m}\vert^2$) were precisely what was needed to
get $C(\theta\geq60^\circ)=0$, cosmic variance would perversely obliterate the carefully adjusted relationship among the $C_\ell$ of the theory
would not be preserved by the measured values on the sky in a particular realization of the $a_{\ell m}$.     To be precise,
when we replace the theoretical $C_\ell$ by the  $C_\ell$ inferred from a Legendre polynomial series expansion of the cut-sky ILC $C(\theta)$,
we find that there is less than a $3\%$ chance of recovering an $S_{1/2}$ less than or equal to the observed value of $S_{1/2}$.
Moreover, most of those $3\%$ achieve the low $S_{1/2}$ by lowering multiple low-$\ell$ $C_\ell$, and so are unlike the
observed angular power spectrum.    The observed sky, at least the part outside the Galaxy cut,  seems not to respect the fundamental
prediction of the standard cosmological model that the $a_{\ell m}$ are independent random variables.

\section{Summary}
    
The inflationary $\Lambda$CDM model has many successes.  
The ability to fit the peaks and troughs of the  medium and high-$\ell$ CMB TT angular power spectrum
with just a few parameters is remarkable.   
However, for the lowest few multipoles and the largest angular scales, the observations disagree markedly with the predictions of the theory.
Examining the lowest interesting multipoles (the quadrupole and octopole) of the best full sky CMB map, 
we find that they appear unexpectedly correlated with each other.  The plane defined by the quadrupole and the three
planes defined by the octopole are nearly parallel to each other.  They are nearly perpendicular to the plane of the Solar System (ecliptic).
They point essentially at the dipole -- the direction of our motion through the CMB.  Finally, they are  oriented (with respect to their shared axis) 
such that the ecliptic carefully separates the strongest extrema in the north from the weaker extrema of the south.   
(Any review of CMB anomalies would include multiple other examples, some of which may well be connected to the above.)

These deviations from statistical isotropy  in our CMB sky have yet to be explained, 
and there are significant challenges to doing so.   
Because of their multipole structure, these deviations are not characteristic of something one would obtain from a mis-handling of the Galactic foregrounds,
and they are also difficult to obtain from a single localized patch of the sky.
They are also not easy to obtain by a perturbative expansion in small vectors (errors in the dipole, gradients of potentials, {\it etc}.) 
because the quadrupole is so much smaller than the octopole.

Arguably one should not trust the part of the sky behind the Galaxy.   
%Masking that out makes it impossible to reconstruct the spherical harmonic components well enough to verify the alignments seen on the full sky.   
%However, 
Examining the two-point angular correlation on the sky outside the Galaxy we find that
there is a marked absence of correlations above $60^\circ$ angular separation.  By one
measure (first proposed by WMAP),  this absence of correlation has a p-value of just
$0.025\%$; in other words. it would happen accidentally in the best-fit $\Lambda$CDM model
just once in 4000 realizations.   Most troublingly, it suggests that  $C_\ell$ of different $\ell$ are
not independent.  

This anomaly too has, so far, found no satisfactory explanation.   One could imagine that non-trivial cosmic topology
could induce covariance among $C_\ell$ (since the fundamental domain does not possess a rotational symmetry),
however searches for non-trivial topology have so far failed to find any \cite{Cornish:2003db,ShapiroKey:2006hm,Bielewicz:2010bh}\footnote{
The circle searches have been extended to non-approximately-antipodal
circle pairs without finding any statistically significant signal \cite{Vaudrevange}
}.
These searches already extend nearly to the diameter of the last scattering surface.
Searches beyond the last scattering surface are possible in principle, but so far none have been demonstrated to be powerful.

Other explanations have been proffered that can reduce $S_{1/2}$ (for example\cite{Afshordi:2008rd}), 
but none that can reduce its expected value to the observed one.
The challenge is how to induce covariance among the $C_\ell$ within the context of the inflationary paradigm.

Future results from the Planck satellite may show these large-angle/low-$\ell$ anomalies to be nothing more than
systematic errors in the measurements or analysis of the WMAP (and the COBE) team, but unless and until they do
these anomalies remain the outstanding point of disagreement between the standard cosmological model
and observations.

\begin{acknowledgement}
The authors acknowledge the WMAP Science Team for (among many other things) use of multiple figures.
GDS thanks the organizers of the Southeastern European Network in Mathematical and Theoretical Physics for their kind hospitality.
GDS and CJC are supported by a grant from the US Department of Energy to the particle astrophysics theory group at CWRU.
DH is supported by DOE OJI grant under contract DE- FG02-95ER40899, and NSF under contract AST-0807564. 
DH and CJC are supported by NASA under contract NNX09AC89G. 
This research was also supported in part by the NSF Grant No. NSF PHY05-51164.
DJS is supported by Deutsche Forschungsgemeinschaft (DFG). 
This work made extensive use of the HEALPIX package \cite{Gorski:2004by}.
The numerical simulations were performed on the facilities provided by the Case ITS High Performance Computing Cluster.
\end{acknowledgement}

\end{document}